\documentclass[12pt]{article}
\usepackage{color}
\usepackage{amsfonts}
\usepackage{graphicx}
\usepackage{epsfig}
\usepackage{amsmath}
\usepackage{amssymb}
\usepackage{multirow}
\usepackage{cancel}
\usepackage{hyperref}
\begin{document}
\begin{titlepage}
\begin{flushright}
RECAPP-HRI-2013-017\\
SINP/TNP/2013/09
\end{flushright}
\vspace{.5cm}

\begin{center}
{\Large {\bf 
{Drell-Yan, ZZ, W$^+$W$^-$ production in SM \& ADD model to NLO$+$PS accuracy at the 
LHC }}} 
\\[1cm]

R.\ Frederix$^{a}$,
M.\ K.\ Mandal$^{b}$,
P.\ Mathews$^{c}$,
V. Ravindran$^{d}$,
S.\ Seth$^{c}$
\\[1cm]

${}^a$ 
PH Department, TH Unit, CERN, CH-1211 Geneva 23, Switzerland
\\[.5cm]

${}^b$ Regional Centre for Accelerator-based Particle Physics\\ 
Harish-Chandra Research Institute, Chhatnag Road, Jhunsi,\\
Allahabad 211 019, India
\\[.5cm]

${}^c$ Saha Institute of Nuclear Physics, 1/AF Bidhan Nagar, Kolkata 700 064, 
India
\\[.5cm]

${}^d$
The Institute of Mathematical Sciences\\
C.I.T Campus, 4th Cross St, Tharamani  Chennai,\\
Tamil Nadu 600 113, India
\\[.5cm]

\end{center}
\vspace{1cm}

\begin{abstract}
\noindent 
In this paper, we present the next-to-leading order QCD corrections for di-lepton,
di-electroweak boson ($ZZ$, $W^+W^-$) production in both the SM and the ADD model,
matched to the HERWIG parton-shower using the {\scriptsize A}MC@NLO framework.  A
selection of results at the 8 TeV LHC, which exhibits deviation from the SM as a
result of the large extra-dimension scenario are presented.
\end{abstract}

\vspace{.7cm}

\end{titlepage}

\section{Introduction}

With more accumulated data at the LHC, extra dimension searches at different
energies have yielded stringent bounds \cite{cms,atlas} on the model parameters
\cite{ADD,RS1}.  This has also been facilitated by improved theoretical
calculations to next-to-leading order (NLO) in QCD that have been available for the
large (ADD) \cite{ADD} and warped (RS) \cite{RS1} extra dimension models for
various processes {\em viz.}\ di-lepton \cite{dy}, di-boson ($\gamma \gamma$
\cite{diph}, $ZZ$ \cite{diZ}, $WW$ \cite{diW} ($W^+ W^-$ is denoted as $WW$)).
In extra dimension models, pair production could result from the exchange of
virtual Kaluza-Klein (KK) modes.  As a result of possible new physics, it is
expected that the production rate and potentially certain kinematical
distributions may get modified as compared to the SM predictions.  Further,
it is essential that higher order QCD corrections are included as it leads to
reduction in scale uncertainties which in turn improves the theoretical
predictions.  For extra dimension searches, ATLAS and CMS have investigated
the impact of NLO corrections in their analysis by using constant K-factors,
which does not necessarily give reliable predictions.

One important recent development has been the implementation of the di-photon
production to NLO including Parton Shower (PS) in the {\scriptsize A}MC@NLO
environment for the ADD model \cite{dprecent}.  This allows for the generation
of fully exclusive events that are NLO accurate for observables inclusive in
QCD radiation.  If required, these events can be directly passed through a
detector simulation.  In this paper, we have implemented the rest of the pair
production processes ($\ell^+ \ell^-$, $ZZ$ and $WW$) that could contribute 
to the ADD model, to NLO+PS accuracy in the {\scriptsize A}MC@NLO environment.  

To set the notations and the conventions used, we briefly describe the
interaction Lagrangian 
\begin{eqnarray}
{\cal L}_{int} = - \frac{\kappa}{2} \sum_{(\vec n)} T^{\mu \nu} 
h_{\mu \nu}^{(\vec n)} ~,
\end{eqnarray}
of the massive spin-2 KK modes $h_{\mu\nu}^{(\vec n)}$ with the SM particles,
which is through the energy momentum tensor $T^{\mu\nu}$ of the SM.  The coupling
$\kappa$ is related to the Planck mass in 4-dimension, $\kappa=\sqrt{16 \pi} /M_P$.
Using the convention of HLZ \cite{HLZ} the summation of the KK modes in the 
propagator ${\cal D} (s)$ is given by
\begin{eqnarray}
\kappa^2 {\cal D} (s) &=& \kappa^2 \sum_{n} \frac{1}{s-m_n^2+i 
\epsilon} ,
\nonumber \\
&=& \frac{8 \pi}{M_S^4} \left( \frac{\sqrt{s}}{M_S}\right )^{(d-2)}
\left[ -i \pi + 2 I \left(\frac{\Lambda}{\sqrt{s}}\right) \right] .
\label{KKsum}
\end{eqnarray}
The summation over KK modes leads to the integral $I(\Lambda/\sqrt{s})$,
defined in \cite{HLZ}, $\sqrt{s}$ is the center of mass energy, $\Lambda$
is the UV cutoff of the KK modes which is identified with the fundamental
scale $M_S$ in $4+d$ dimensions \cite{HLZ,GRW}.   Bounds on $M_S$ for
different extra dimensions $d$ have been obtained by ATLAS and CMS
collaborations; for our present analysis we choose the following values
$M_S=$ 3.7 TeV (d=2), 3.8 TeV (d=3), 3.2 TeV (d=4), 2.9 TeV (d=5), 2.7 TeV
(d=6). 

The rest of the paper is as follows: we briefly describe the framework for
matching the NLO results with Parton Shower Monte Carlo in section 2.
A selection of the numerical results are presented in section 3
and finally we present our conclusions in section 4.

\section{NLO+PS}

In order to provide a more realistic description of a process at the LHC, it is
unavoidable to match the NLO QCD results with Parton Shower Monte Carlo. For
the present analysis, we adopt the MC@NLO formalism \cite{mcnlo} to match the
fixed order NLO results with the HERWIG6 \cite{herwig} parton shower, including
the hadronisation contribution by using the automated {\scriptsize A}MC@NLO
framework. The Born and real-emission correction for all these processes are
computed with M{\scriptsize AD}FKS \cite{madfks}, which uses the FKS
subtraction method \cite{fks} to compute the real-emission contribution in an
automated way, within the MadGraph5 \cite{mg5} environment.  The virtual 
contributions are implemented separately in this environment for each of these 
processes, using the analytically calculated results for $\ell^+ \ell^-$ \cite{dy},
 ZZ \cite{diZ} and WW \cite{diW} production processes. 
We have also incorporated an algorithm that takes care of the summation of the
KK modes in the ADD model (Eq.\ \ref{KKsum}); this has been made possible by
appropriate changes in the spin-2 HELAS routine \cite{dprecent}. 
The exact numerical
cancellations of double and single poles coming from the real and virtual terms
in all the subprocesses, for each of the production processes have been checked.

For the Drell-Yan (DY) process, we have generated the events for the process $PP\rightarrow
e^+e^- ~X$, which is phenomenologically same as $P P \rightarrow \mu^+ \mu^- ~X$,
except for the experimental identification of the final state particles.  The
leading order (LO) partonic contribution comes from
the $q\ \bar{q}\rightarrow e^+e^-$ in both the SM and ADD model, whereas
at LO $g\ g\rightarrow e^+ e^-$ contributes only to the ADD model.  Emission of
real gluon and one loop correction due to the virtual gluon, together with the
partonic subprocess $q(\bar{q})\ g\rightarrow q(\bar{q})\ e^+e^-$,
give all the ${\cal{O}}(\alpha_s)$ contributions.
The interference between the SM and ADD diagrams also give ${\cal{O}} (\alpha_s)$
contribution at the NLO.  For the di-boson final states, in addition to similar
partonic sub processes, there are contributions due to the interference
between the $gg$ initiated box diagrams in SM and the $gg$ initiated Born diagrams
in the ADD which is of ${\cal{O}}(\alpha_s)$.  We have considered all the above
contributions in each of these processes of interest for our present analysis.

After generation of events following the above procedure,
we let the $Z$ and $W^\pm$ bosons to decay to leptons at the time of showering.
For the $ZZ$ events, we let one $Z$ boson to decay to $e^+ e^-$ and the other 
one to $\mu^+\mu^-$, while for $WW$ events we let the $W^+$ decay to $e^+ \nu_e$
and the $W^-$ to $\mu^- \bar \nu_\mu$.
Alternatively, the $W^\pm$ and $Z$ bosons can be decayed using MadSpin \cite{MadSpin}
at the time of event generation itself,
which retains nearly all spin correlations.  We have not chosen to do this, because
the inclusion of the sum over the KK modes is non-trivial in this way.

\section{Numerical Result}

In this section, we present some of the kinematical distributions for the
production of $\ell^+ \ell^-$, $ZZ$, $WW$, both in the SM and ADD to NLO+PS
accuracy for the LHC center of mass energy $\sqrt{S}=8$ TeV.  Events are
generated using the following input parameters: $\alpha_{EW}^{-1} = 132.507$,
$G_F = 1.16639\times10^{-5}$ GeV$^{-2}$, $m_z = 91.188$ GeV.  Using these
electro-weak parameters as inputs, the mass of $W$ boson $m_w = 80.419$ GeV
and $\sin^2\theta_w = 0.222$ are obtained.  The (N)LO events are generated
using MSTW(n)lo2008cl68 parton distribution functions (PDF) for the (N)LO 
and the value of strong coupling constant $\alpha_s$ is solely determined
by the corresponding MSTW PDF \cite{mstw} at (N)LO. The factorisation scale
$\mu_F$ and the renormalisation scale $\mu_R$ are set equal to the invariant
mass of the corresponding di-final state.  The number of active quark flavor
is taken to be five and are treated as massless.  We use the following loose
cuts at the time of event generation for the DY production: (a) transverse
 momentum of the lepton $P_T^{\ell} > 15$ GeV, (b) rapidity 
$|\eta^{\ell}| < 2.7$, (c) the separation of two particles in the 
rapidity-azimuthal angle plane $\Delta R^{e^+e^-} > 0.3 $ (where 
$\Delta R = \sqrt{(\Delta \eta)^2 + (\Delta \phi)^2}$) and (d) the invariant
mass $ M_{e^+e^-} < 1.1\times M_S$. For $ZZ$ and $WW$ event generation, we use
no cut at the generation level except on the invariant mass {\em i.e.},
$M_{ZZ},M_{W^+W^-} < 1.1\times M_S$.  For $WW$ event generation, the following
CKM matrix elements are used: 
$|V_{ud}| = 0.97425$, $|V_{us}| = 0.2252$, $|V_{ub}| = 4.15\times10^{-3}$, 
$|V_{cd}| = 0.230$, $|V_{cs}| = 1.006$, $|V_{cb}| = 40.9\times10^{-3}$.  All
the CKM matrix elements associated with the top quark are taken to be zero.
\begin{figure}[tbh]
\centerline{
\includegraphics[width=5cm]{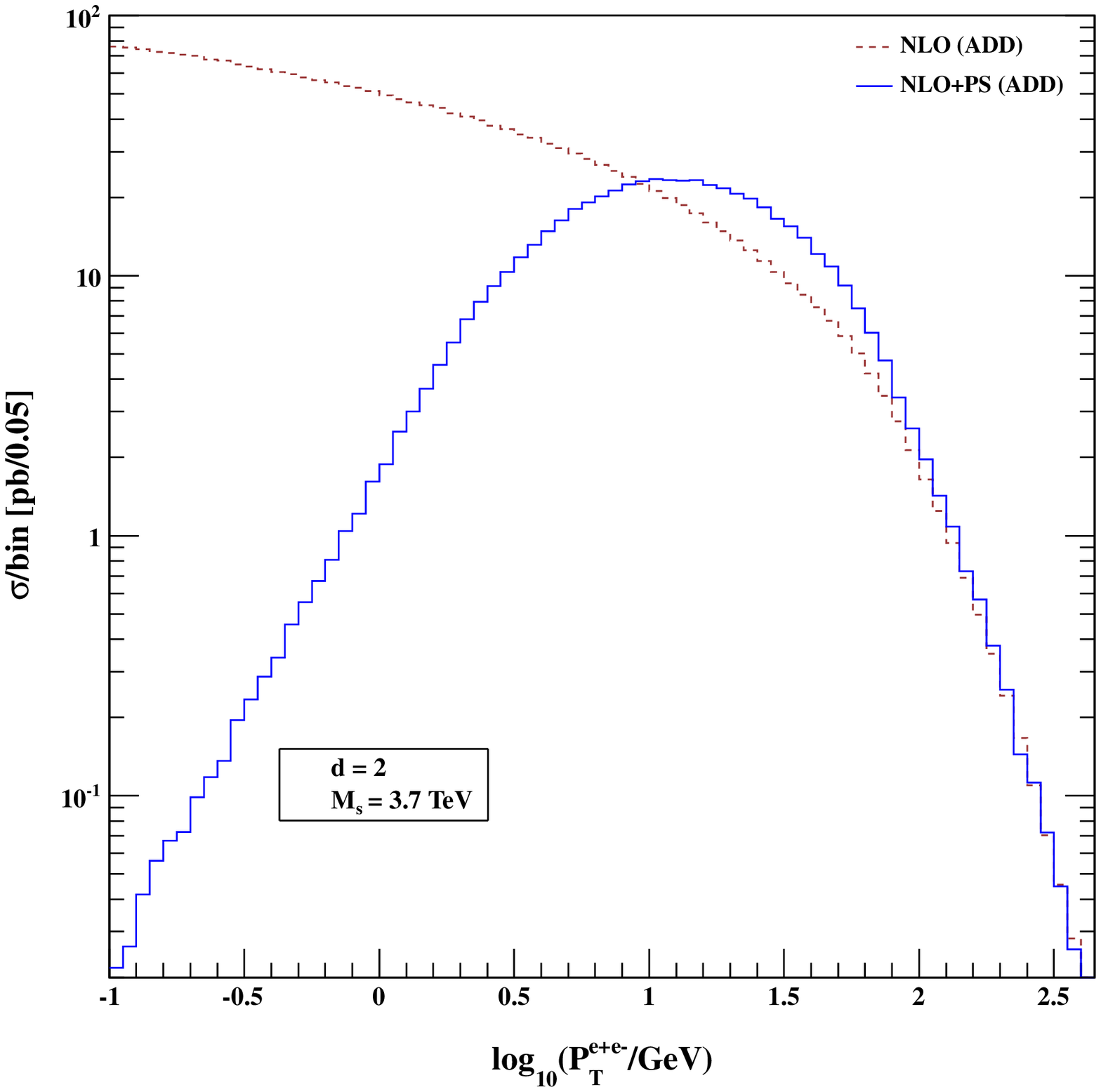}
\includegraphics[width=5cm]{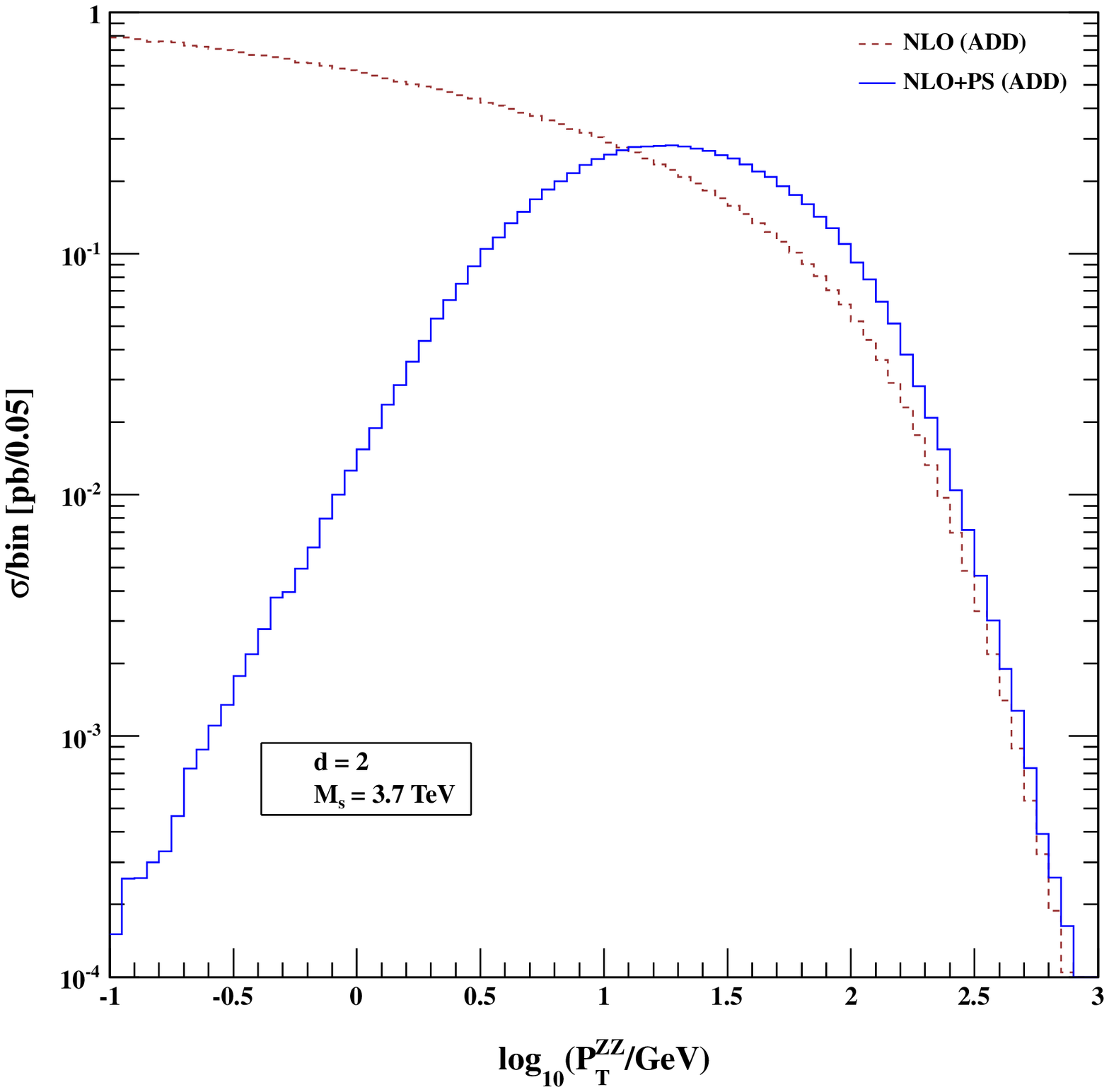}
\includegraphics[width=5cm]{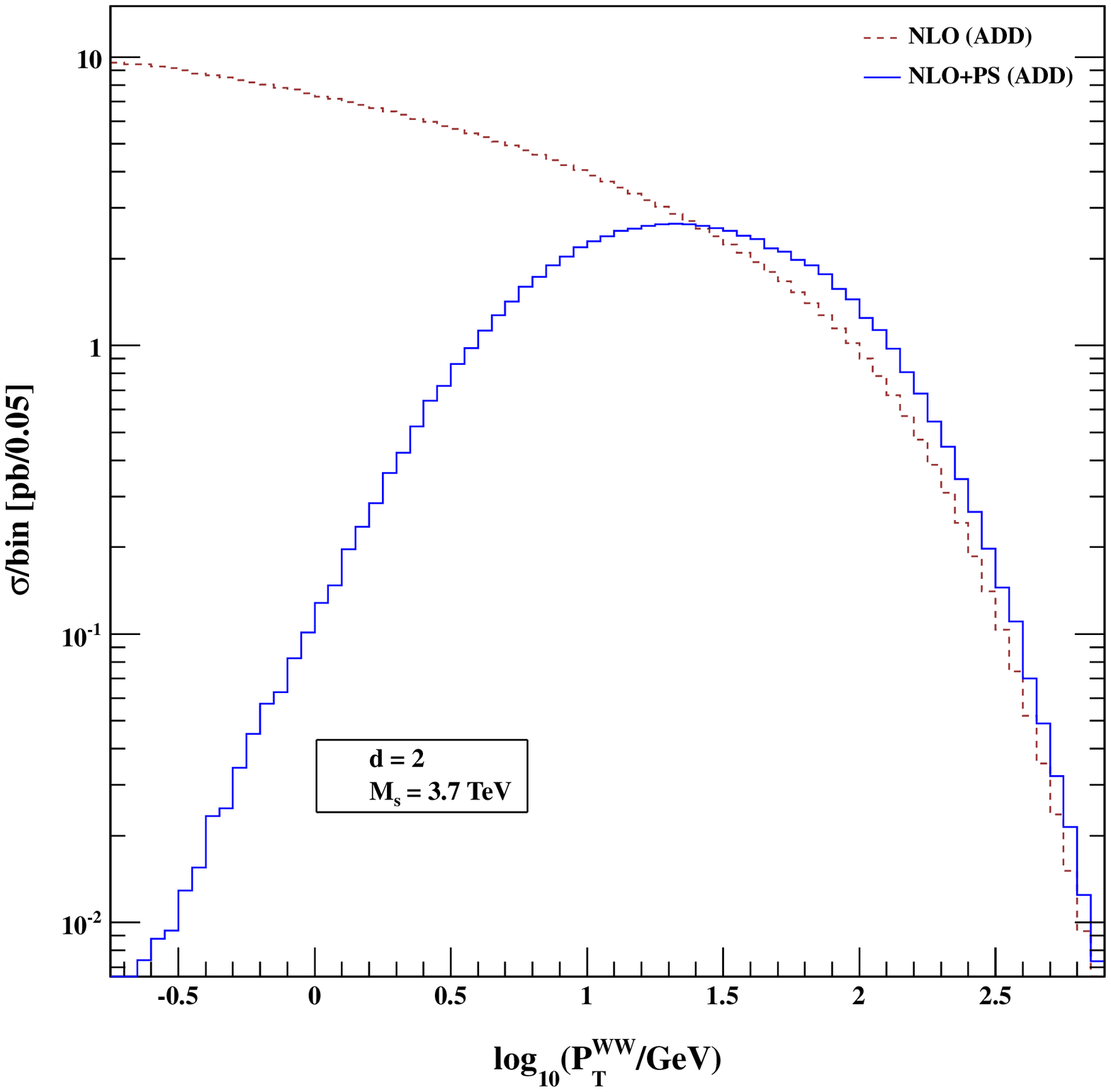}
}
\caption{\label{fo}
Fixed order NLO results (dashed brown) along with the NLO+PS results (solid blue)
for the $\log_{10}(P_T)$ distribution of the $e^+e^-$ (left), $ZZ$ (middle) and
$W^+W^-$ (right) pair.}
\end{figure}

For showering the DY events, HERWIG6 in MC@NLO formalism is used.  Using the
following analysis cuts: $P_T^{l} > 20$ GeV ($l = e^+, e^-$), $|\eta^{l}| < 2.5$,
$M_{e^+e^-}<M_S$, $\Delta R^{ll} > 0.4 $ for showering, the hardest (with maximum
$P_T$) $e^+$ and $e^-$ are collected.  In order to separate leptons from jets, 
$\Delta R^{lj} > 0.7$ is used.  For both $ZZ$ and $WW$ showering, we have identified
those final state, stable lepton-pair, whose mother is one of the $Z$ boson (for 
$ZZ$ showering) or the final state stable lepton-neutrino pair whose mother is one of the
$W$ boson (for $WW$ showering) and that is the reason we avoid the cut which is
commonly used to reconstruct the $Z$($W$) boson mass from the invariant mass of
the lepton-lepton (lepton-neutrino) pair.  For decay products of $Z/W$, we use
the same analysis cuts to plot various differential distributions and they are
the following: invariant mass $M_{ZZ/W^+W^-} < M_S$, $P_T^{l} > 20$ GeV (where,
$l = e^+, e^-, \mu^+, \mu^-$ for $ZZ$ and $l = e^+, \mu^-$ for $W^+W^-$),
$|\eta^{l}| < 2.5$.  In addition, we have collected only those leptons whose
separation from other leptons and jets are greater than 0.4 and 0.7 respectively
in the rapidity-azimuthal angle plane.
\begin{figure}
\centerline{
\includegraphics[width=8cm]{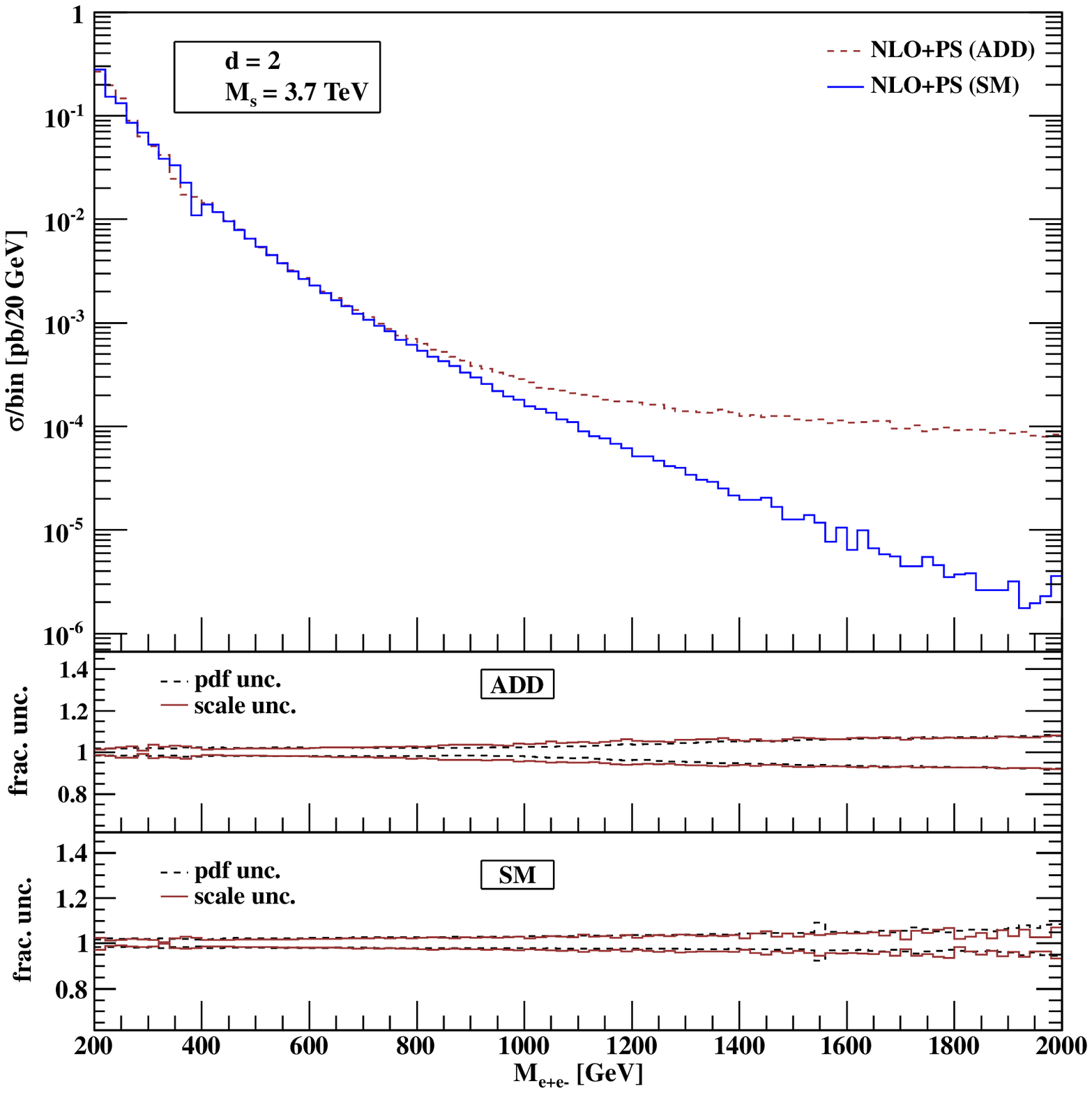}
\includegraphics[width=8cm]{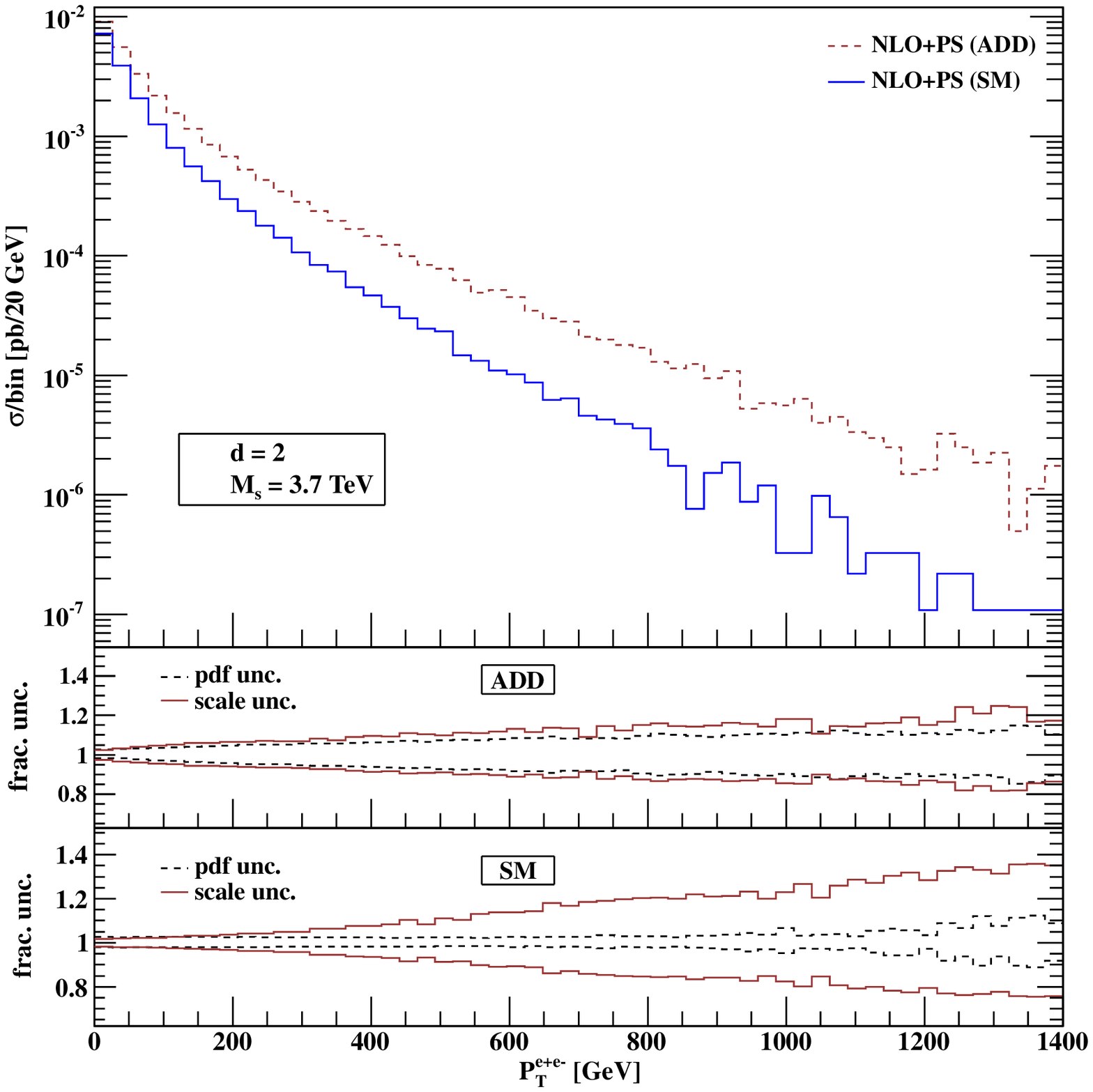}
}
\caption{\label{dy_inv_pt}
Invariant mass ($M_{e^+e^-}$) distribution (left) and transverse momentum distribution
(right) of the $e^+e^-$ pair for ADD ($d=2$) and SM in Drell-Yan process. The right one is
in $M_{e^+e^-} > 600$ GeV region.
}
\end{figure}

Here, we describe few selected differential distributions for some of the
kinematical observables.  To start with, we study the effect of parton shower
over the fixed order NLO correction.  Fixed order NLO results (dashed brown)
along with the NLO+PS results (solid blue) for the $\log_{10}(P_T)$ distribution
of the $e^+e^-$ (left), $ZZ$ (middle) and $WW$ (right) pair are plotted in 
fig.\ \ref{fo}, using their specific analysis cuts detailed above for extra
dimensions $d=2$ and its corresponding $M_S$ value.  In all these plots, the
fixed order cross section diverges for $P_T\rightarrow 0$, while the NLO+PS
result shows a converging behavior in the low $P_T$ region.  The effect of
parton shower ensures correct resummation of the Sudakov logarithmic terms
which appear in the collinear region leading to a suppression of the cross
section in the low $P_T$ region.  There is no significant deviation in the
high $P_T$ region as expected.
\begin{figure}
\centerline{
\includegraphics[width=8cm]{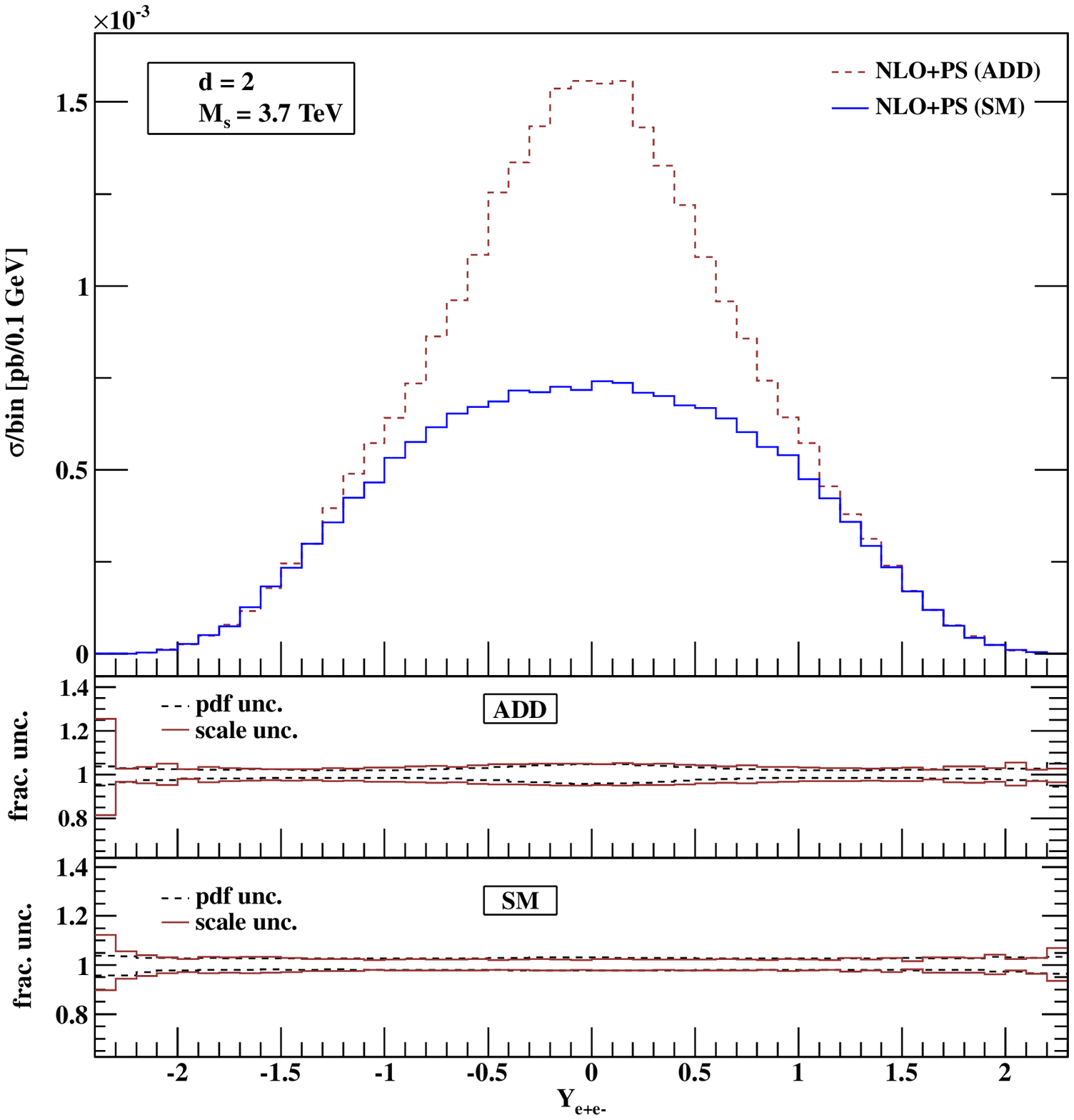}
\includegraphics[width=8cm]{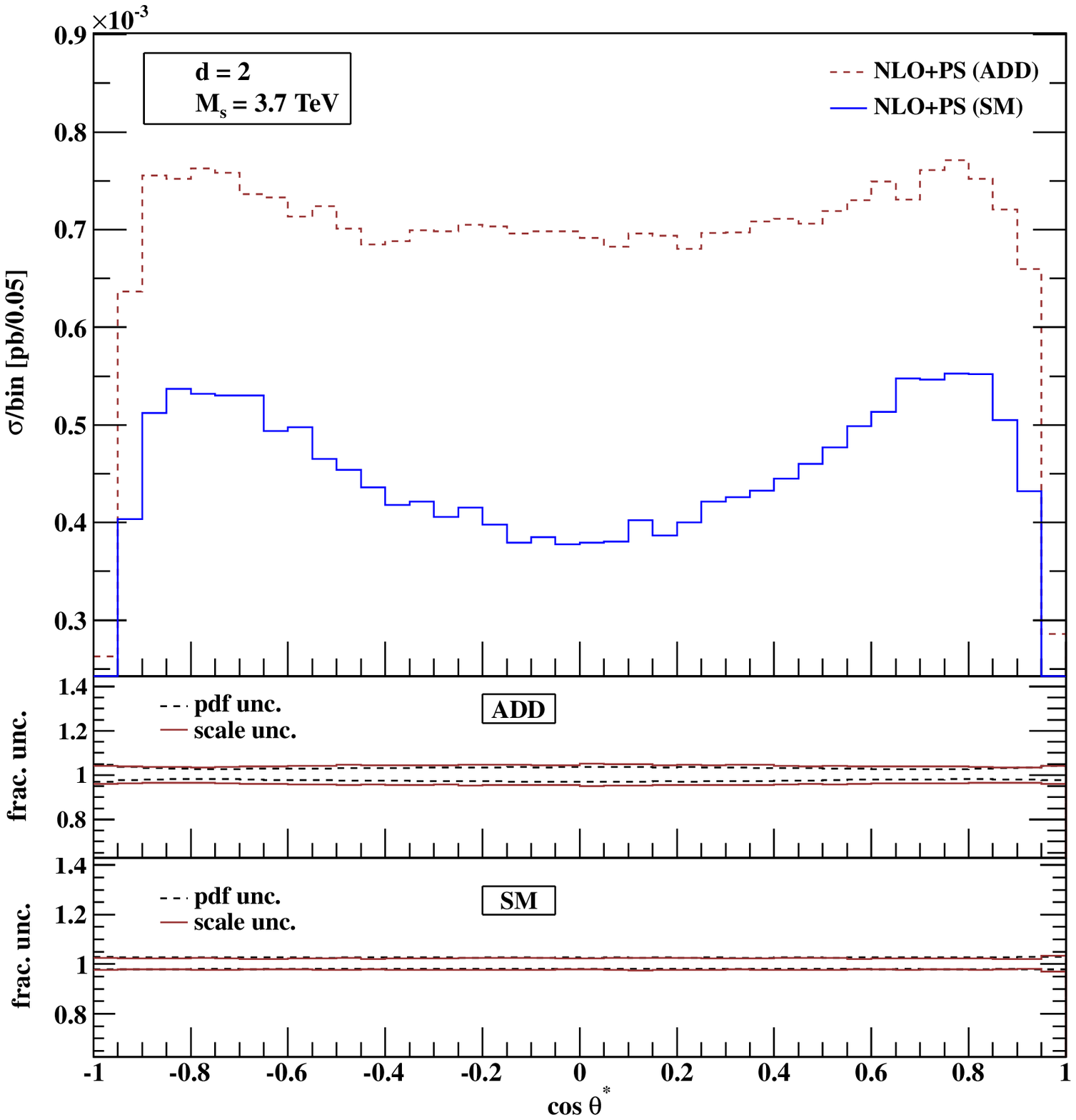}
}
\caption{\label{dy_rap_cos}
Rapidity distribution (left) of $e^+e^-$ pair and the angular distribution
(right) are given for $d=2$ in ADD and also for SM in Drell-Yan process for
$M_{e^+e^-} > 600$ GeV.
}
\end{figure}

In the subsequent plots, we have included fractional scale and PDF uncertainties
corresponding to the SM and ADD model distributions. By fractional uncertainty we
mean the central value of a particular distribution divided by its extremum value.
The scale uncertainties are calculated by considering independent variation of the
renormalisation and the factorisation scales in the following way: $\mu_R=\xi_R M$
and $\mu_F=\xi_F M$. Here, $M$ denotes the invariant mass of the di-final state 
{\em i.e.}, $M_{e^+e^-},M_{ZZ},M_{WW}$ as required and $\xi_R, \xi_F$
can take either of the following values ($1,1/2,2$) independently. The scale
uncertainty band is the envelope of the following $(\xi_F,\xi_R)$ combinations
\cite{dprecent} as described below: (1,1), (1/2,1/2), (1/2,1), (1,1/2), (1,2), (2,1),
(2,2).  Estimation of the PDF uncertainty is done in the Hessian method 
as prescribed by the MSTW \cite{mstw} collaboration.  All these uncertainties are
determined automatically by following the re-weighting procedure \cite{reweight}
built in {\scriptsize A}MC@NLO which stores sufficient information in the parton
level Les Houches events for this purpose.
\begin{figure}
\centerline{ 
\includegraphics[width=8cm]{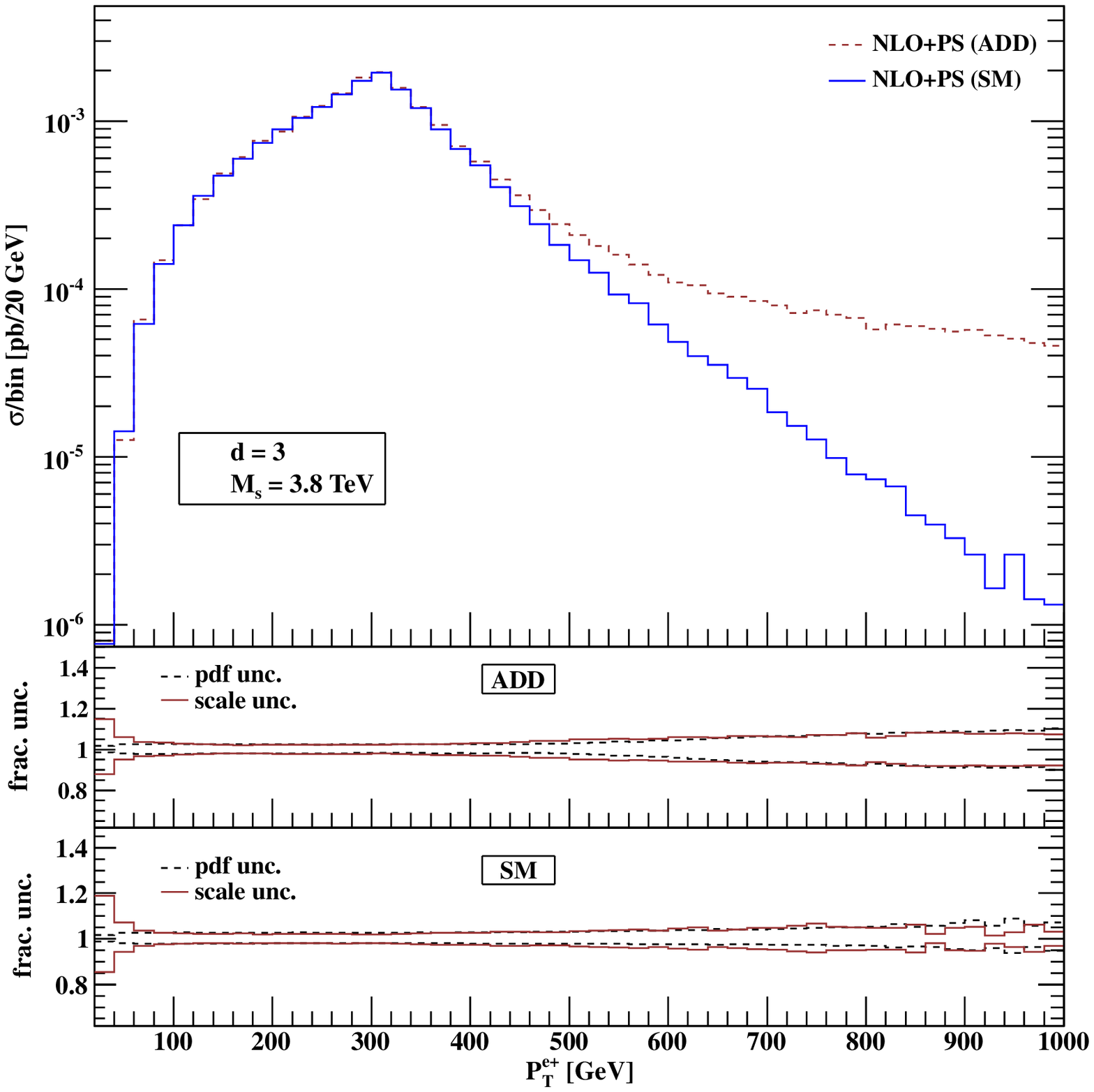}
\includegraphics[width=8cm]{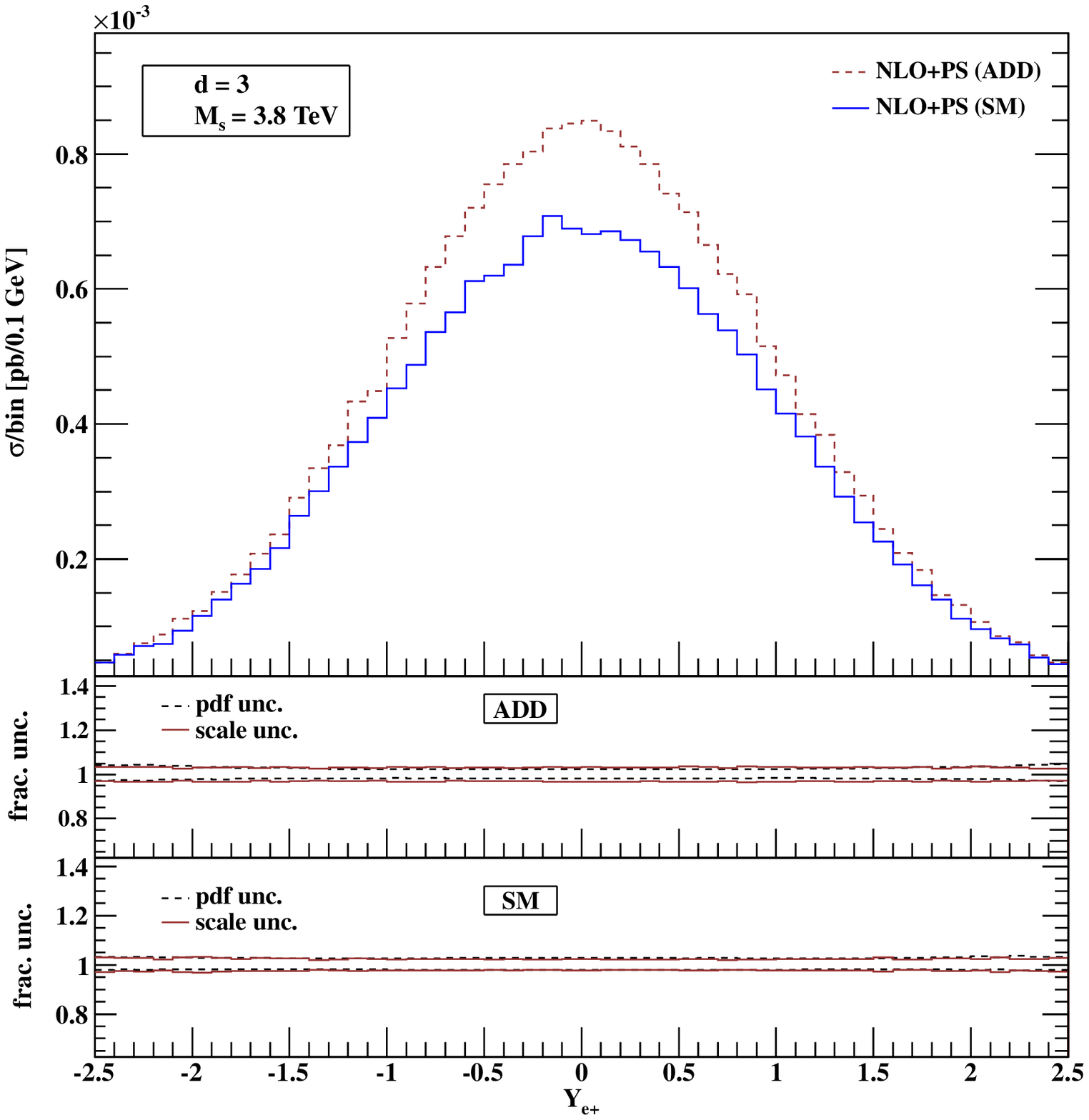}
}
\caption{\label{dy_e2ptrap}
Transverse momentum (left) and rapidity (right) distribution of final state positron
in ADD ($d=3$) and SM for Drell-Yan process for $M_{e^+e^-} > 600$ GeV.
}
\end{figure}

In all the plots ADD represents the full contribution of the SM and 
ADD model contributions including interference.  We use a 
consistent graphical representation for the rest of the kinematic distributions. 
In each case, the upper inset gives the distribution in SM (solid blue) as well as 
in ADD model (dashed brown) to NLO+PS accuracy. For the same distribution, the middle
(ADD) and lower (SM) insets provide fractional scale (solid brown) and PDF (dashed 
black) uncertainties.
\begin{figure}
\centerline{ 
\includegraphics[width=8cm]{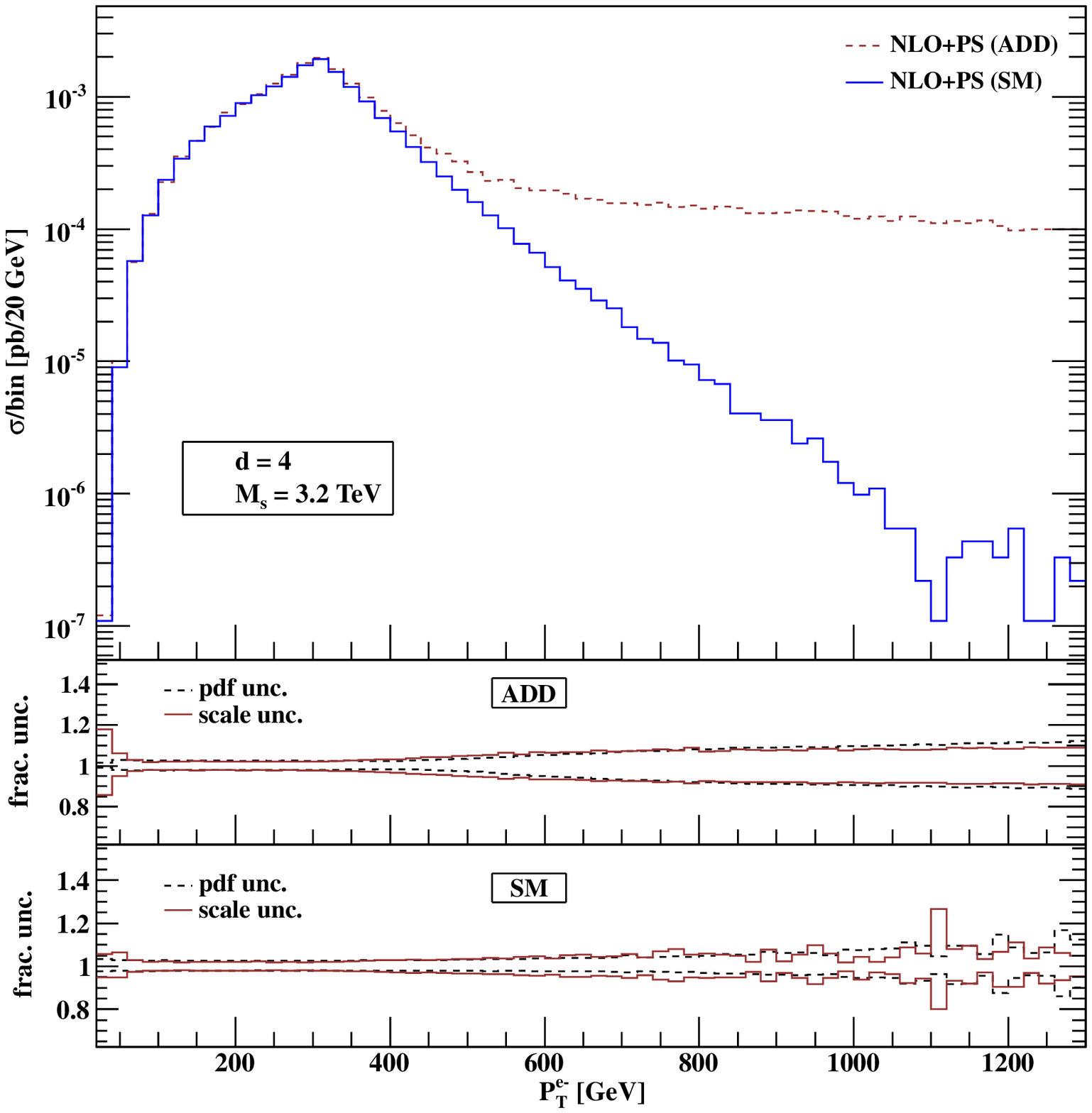}
\includegraphics[width=8cm]{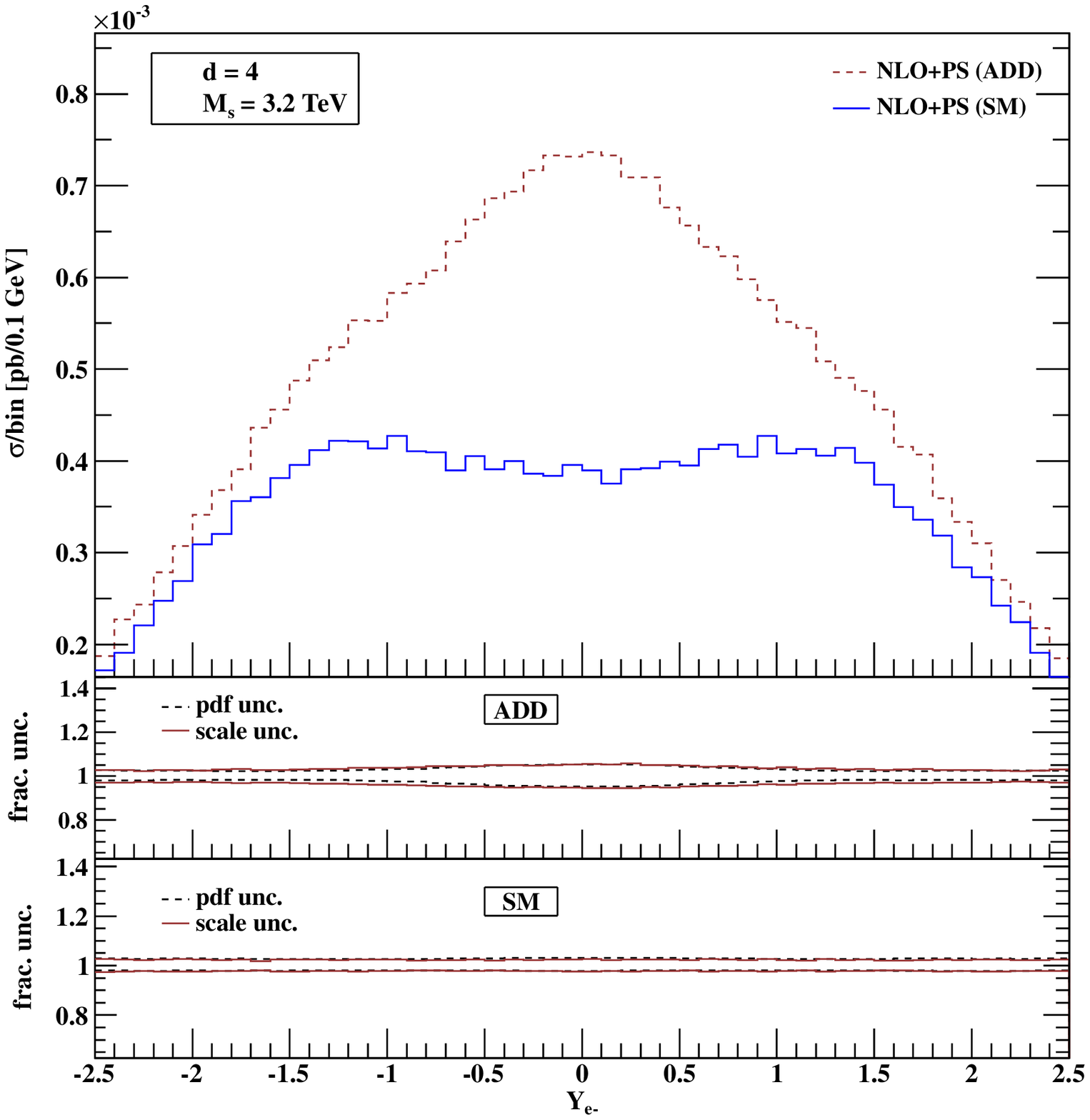}
}
\caption{\label{dy_e1ptrap}
Transverse momentum (left) and rapidity (right) distribution of final state electron
in ADD ($d=4$) and SM for Drell-Yan process for $M_{e^+e^-} > 600$ GeV.
}
\end{figure}

Various kinematical observable in the DY process are given in fig.\ \ref{dy_inv_pt},
\ref{dy_rap_cos}, \ref{dy_e2ptrap} and \ref{dy_e1ptrap}.  In fig.\ \ref{dy_inv_pt},
we have shown the invariant mass distribution (left) and transverse momentum
distribution (right) of the $e^+e^-$ pair for $d=2$ with its associated $M_S$ value.
The effect of large extra dimension is dominant in the high invariant mass region
and hence we focus in the region $M_{e^+e^-} > 600$ GeV to study the other
distribution {\em viz.} $P_T$, rapidity, angular distribution of the $e^+e^-$ pair
and also look at some of the distributions of the individual leptons.  In fig.\
\ref{dy_inv_pt}, note that there is an increase in the scale and PDF uncertainties
with increase in $P_T$ as is well known, see for example \cite{hi_pT}.  In fig.\
\ref{dy_rap_cos}, the rapidity distribution of $e^+e^-$ pair (left) and the
angular distribution (right) are given for $d=2$.  For the rapidity distribution
the deviation from the SM is only prominent in the central region.  The angle made
by the lepton pair in its center of mass frame with respect to one of the incoming
hadron is denoted by $\theta^*$.  The angular distribution is a good
discriminator for the full range to distinguish the ADD from the SM.  fig.\
\ref{dy_e2ptrap} describes the behavior of $P_T$ (left) and rapidity (right)
distribution of final state positron for $d=3$ extra dimensions.  Similarly,
in fig.\ \ref{dy_e1ptrap}, transverse momentum distribution (left) is presented
along with the rapidity distribution (right) of the final state electron for
$d=4$.  The difference in the SM rapidity distribution for $e^-$ (fig.\
\ref{dy_e2ptrap}) compared to $e^+$ (fig.\ \ref{dy_e1ptrap}), can be attributed
to the fact that $Z$ boson couples differently to left and right handed fermions
and the high invariant mass cut used to zoom into the region of interest for the
ADD model, enhances this effect.

\begin{figure}
\centerline{ 
\includegraphics[width=11cm]{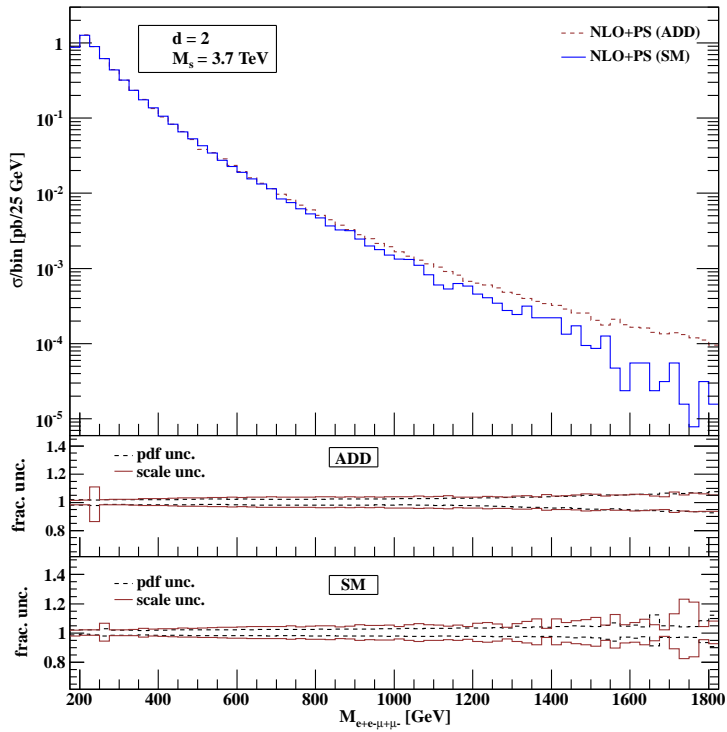}
}
\caption{\label{zz_inv}
Four-lepton invariant mass ($M_{4l}$) distribution for ADD ($d=2$) and SM for decay
products coming from the $ZZ$ process.
}
\end{figure}

\begin{figure}
\centerline{ 
\includegraphics[width=8cm]{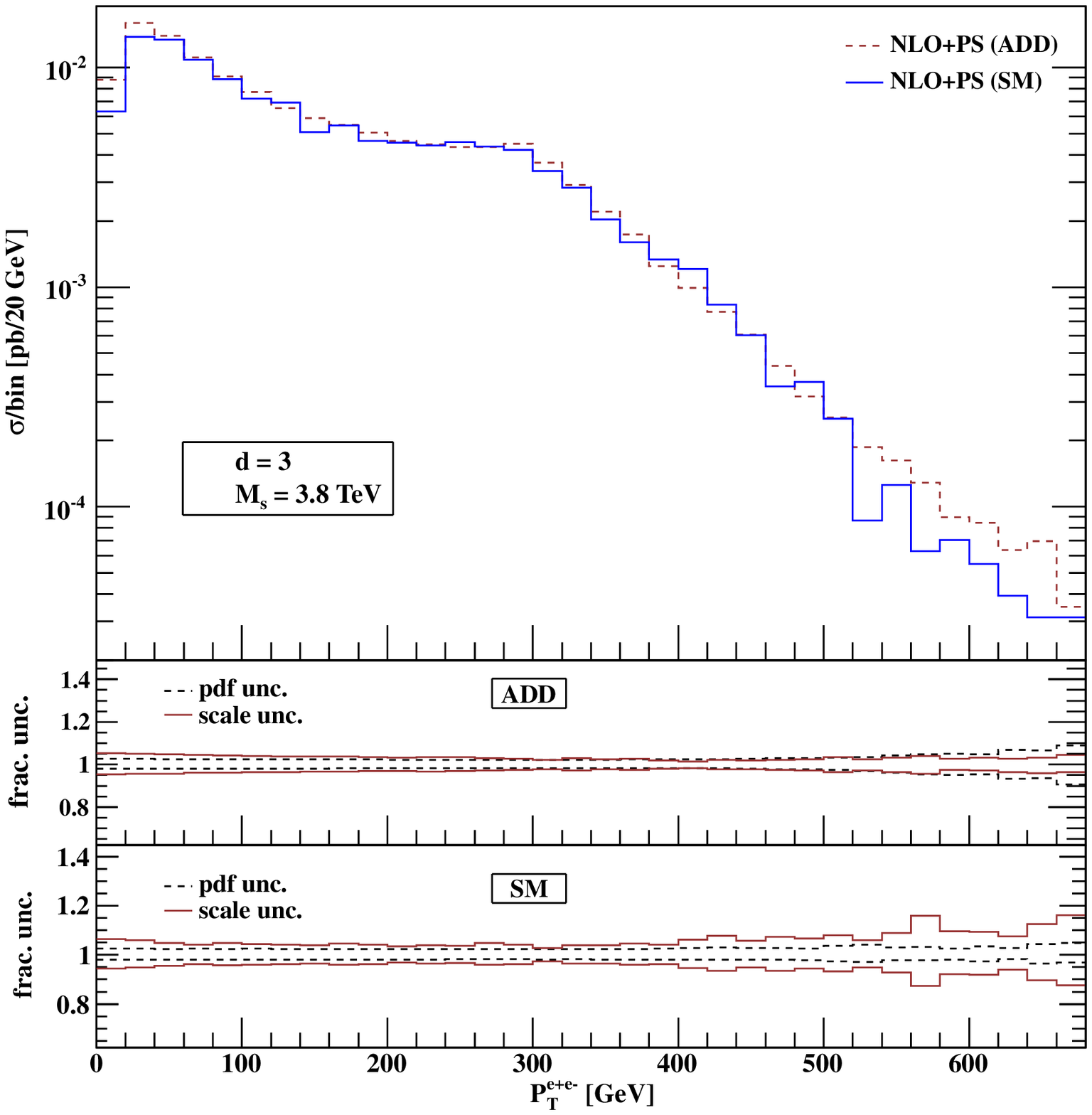}
\includegraphics[width=8cm]{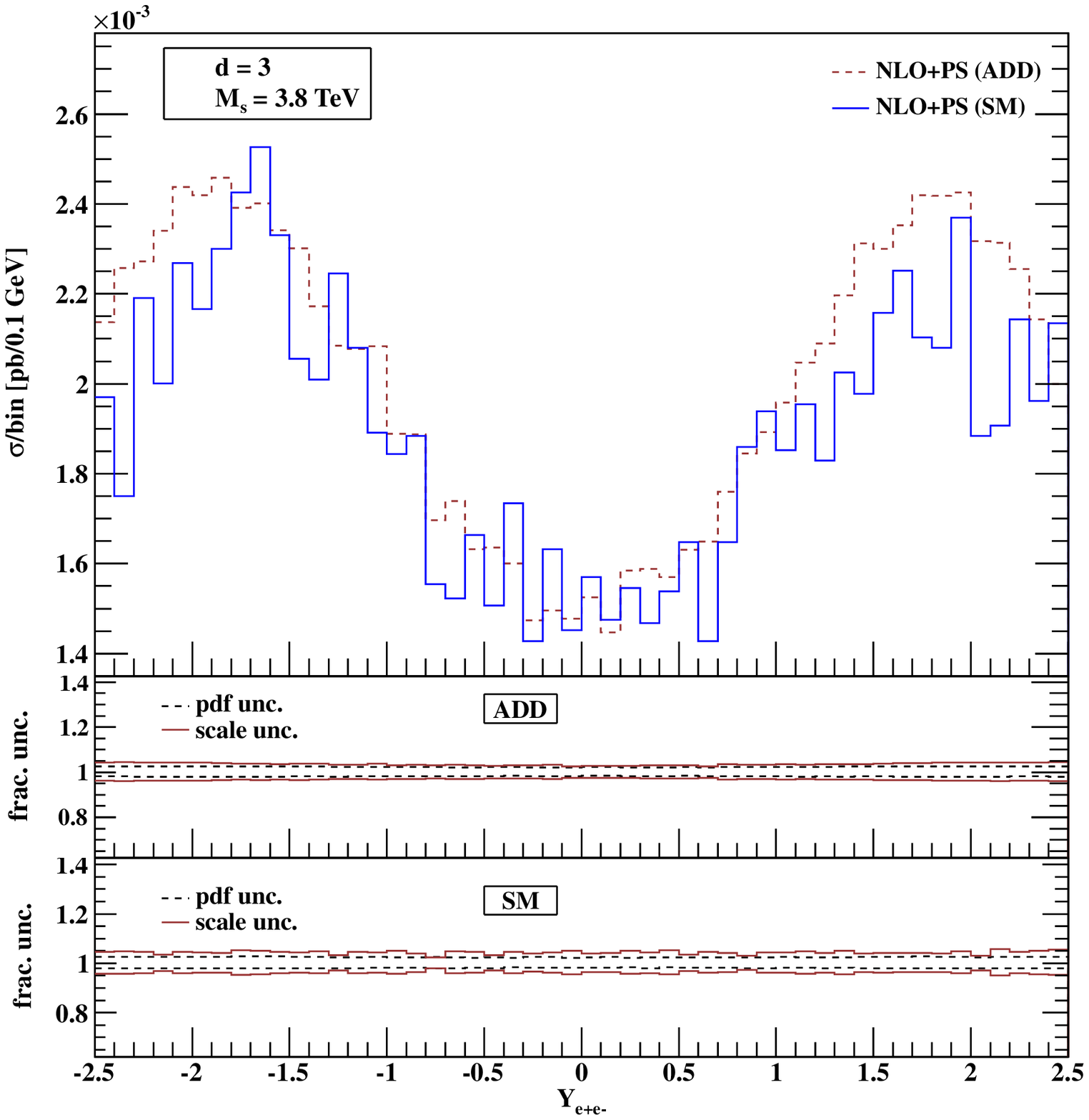}
}
\caption{\label{zz_eeptrap}
Transverse momentum (left) and rapidity (right) distribution of the $e^+e^-$ 
pair coming from $ZZ$ decay for ADD ($d=3$) and SM, when $M_{4l} > 600$ GeV.
}
\end{figure}

\begin{figure}
\centerline{ 
\includegraphics[width=8cm]{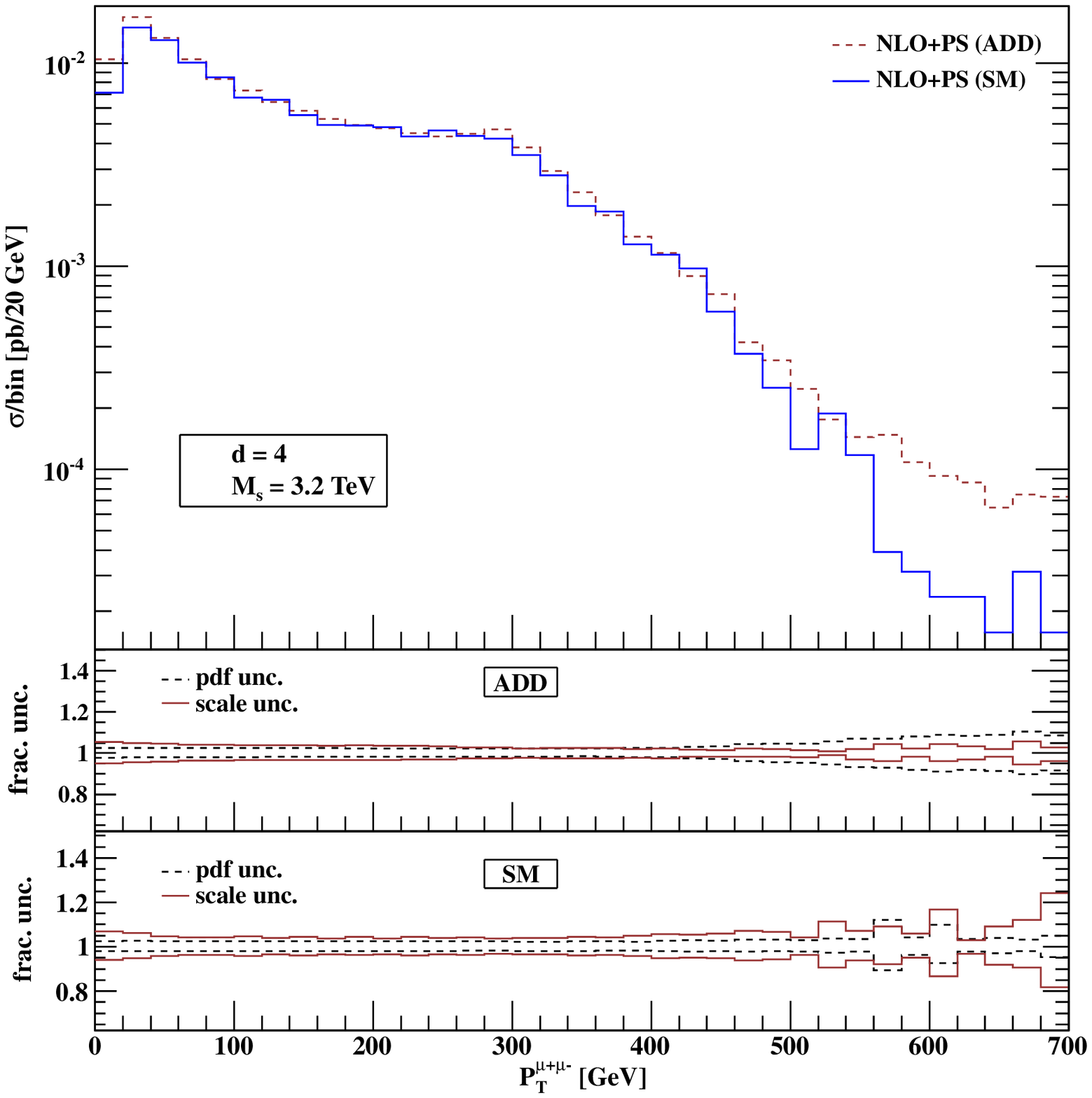}
\includegraphics[width=8cm]{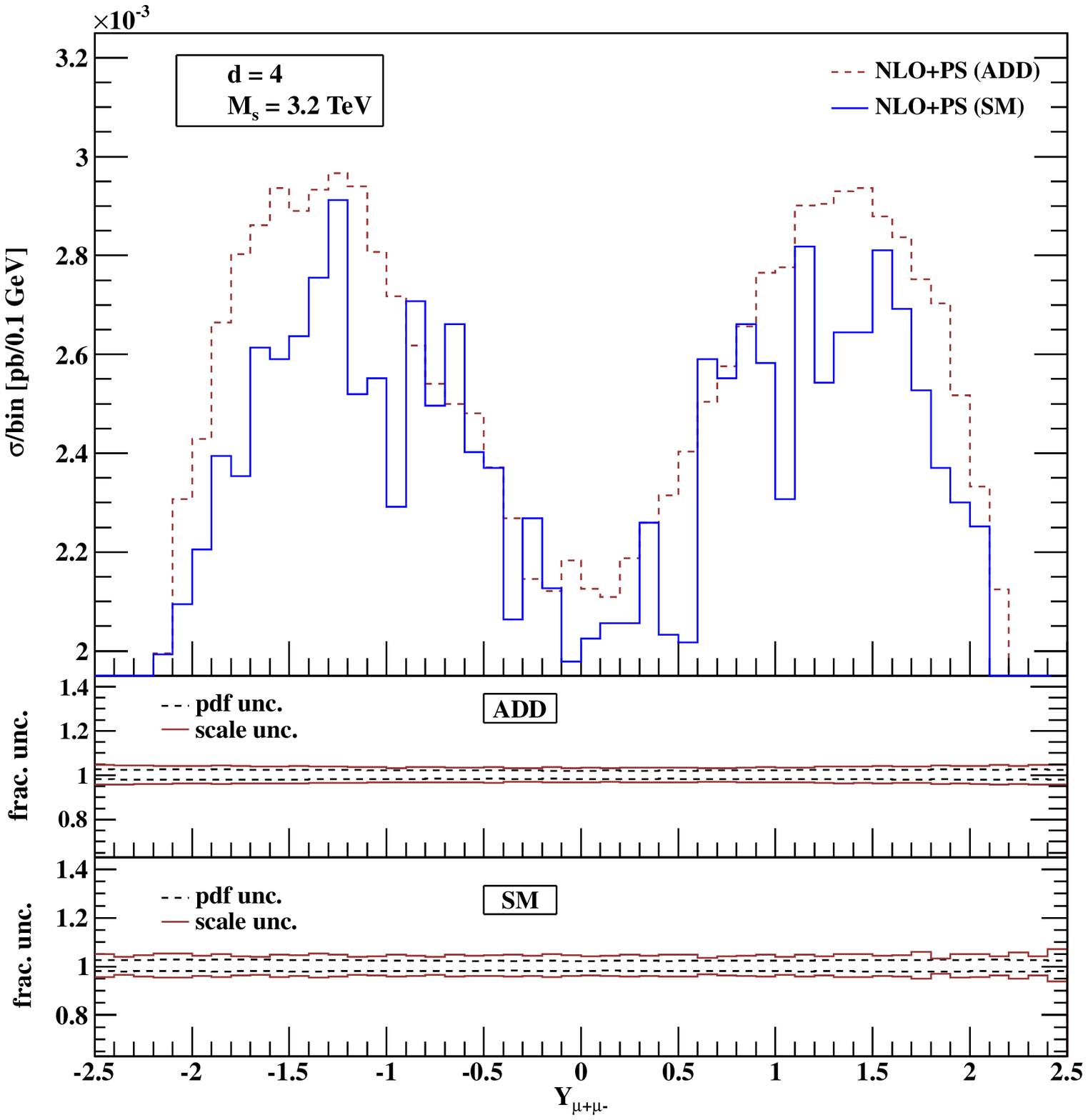}
}
\caption{\label{zz_mmptrap}
Transverse momentum (left) and rapidity (right) distribution of the $\mu^+\mu^-$ 
pair coming from $ZZ$ decay for ADD ($d=4$) and SM, when $M_{4l} > 600$ GeV.
}
\end{figure}

The plots associated with the decay products of $ZZ$ process are presented in
fig.\ \ref{zz_inv}, \ref{zz_eeptrap}, \ref{zz_mmptrap}.  For $d=2$ we see
deviations from the SM in the high invariant mass region in the case of
four-lepton invariant mass ($M_{4l}$) distribution as shown in fig.\ 
\ref{zz_inv}.  Except for the invariant mass distribution, all other kinematical
observable are studied above the region where the four lepton invariant mass is
greater than 600 GeV, which is the ADD dominant region.  In fig.\
\ref{zz_eeptrap}, we have shown transverse momentum (left) and rapidity (right)
distribution of the $e^+e^-$ pair for $d=3$.  Similarly, the transverse momentum
(left) and rapidity (right) distribution for the $\mu^+\mu^-$ pair are presented
in fig.\ \ref{zz_mmptrap} for $d=4$.  The ADD distributions are fairly 
distinguishable for $d=4$ compared to $d=3$, as bounds on $M_S$ value for larger
number of extra dimension is a bit lower. 

\begin{figure}
\centerline{ 
\includegraphics[width=8cm]{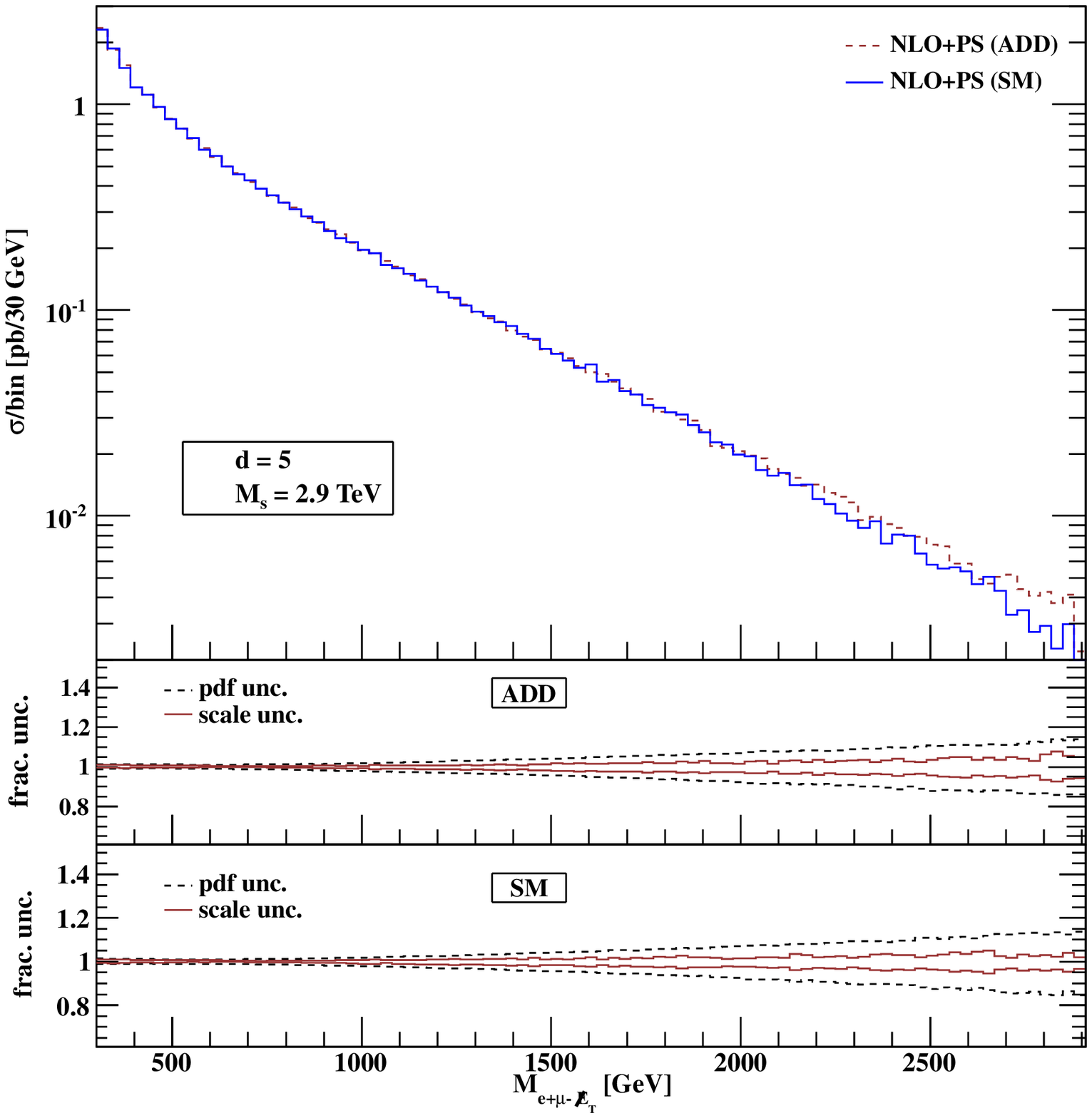}
\includegraphics[width=8cm]{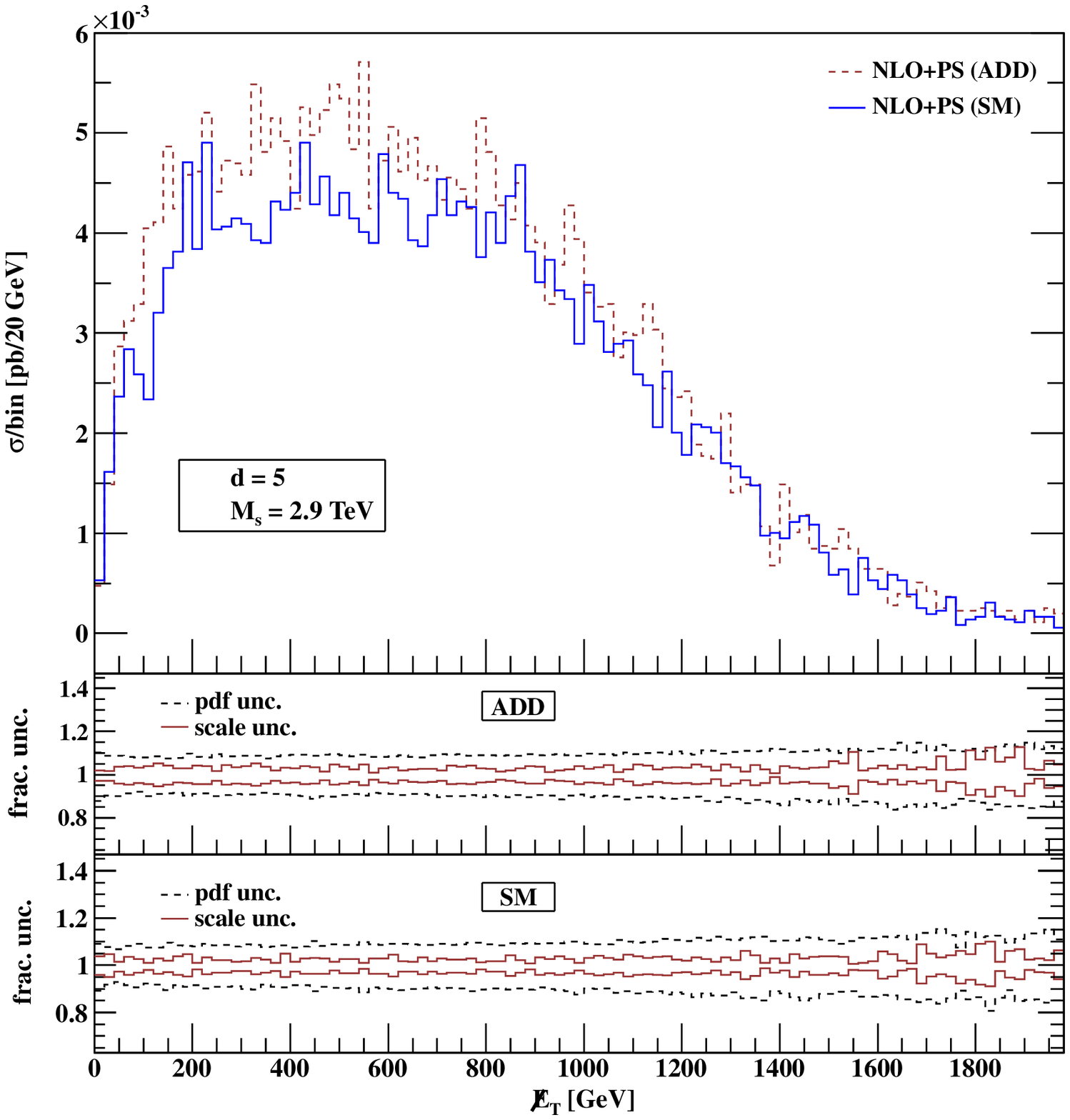}
}
\caption{\label{ww_invmiss}
Invariant mass ($M_{e^+\mu^-\cancel{E_T}}$) distribution (left) of all the final 
state decay products of $WW$ and the total missing transverse energy distribution
(right) for $d=5$ and SM. The right one is restricted within 
$2000 < M_{e^+\mu^-\cancel{E_T}} < M_S$ GeV.}
\end{figure}

\begin{figure}
\centerline{ 
\includegraphics[width=8cm]{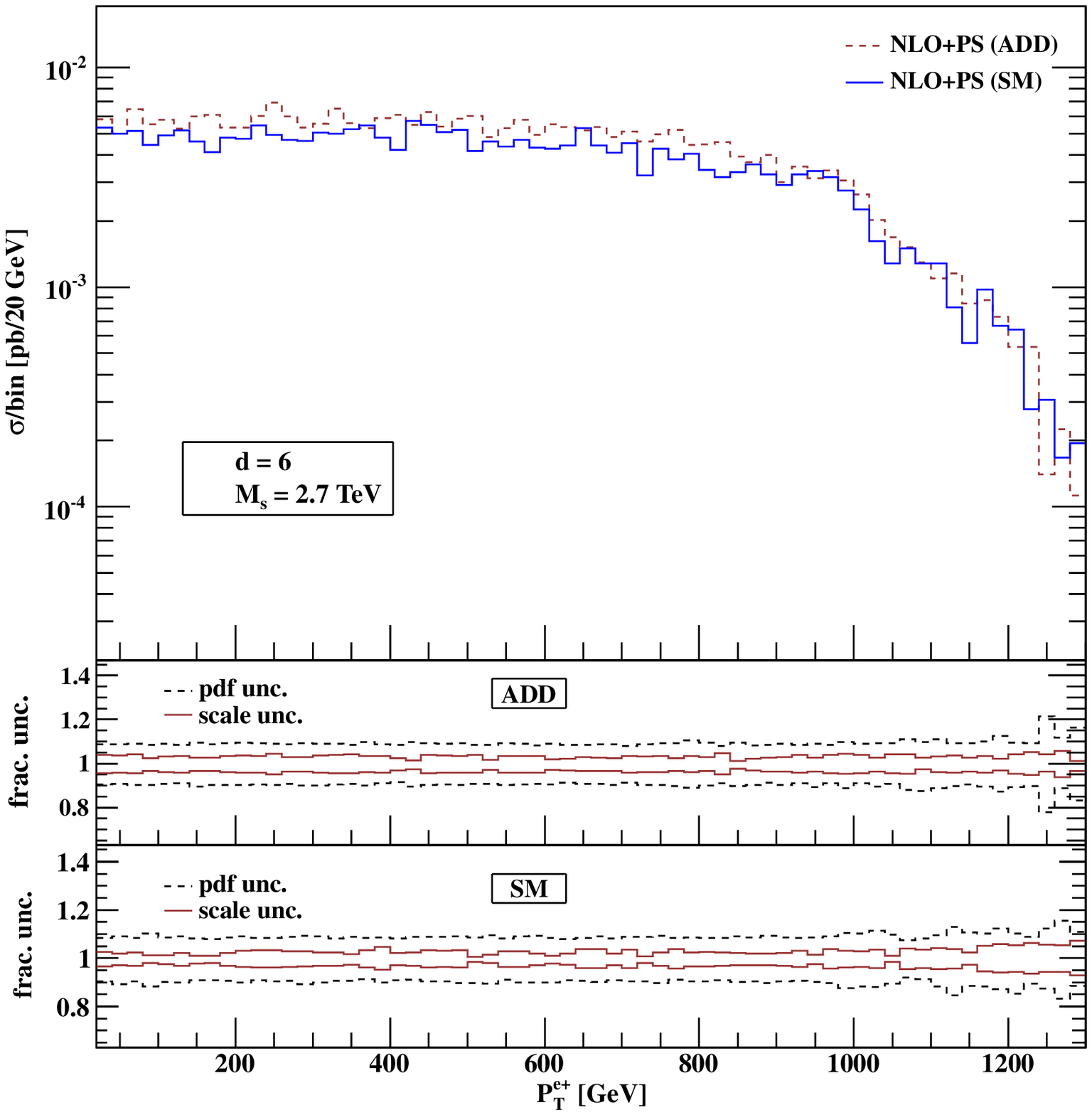}
\includegraphics[width=8cm]{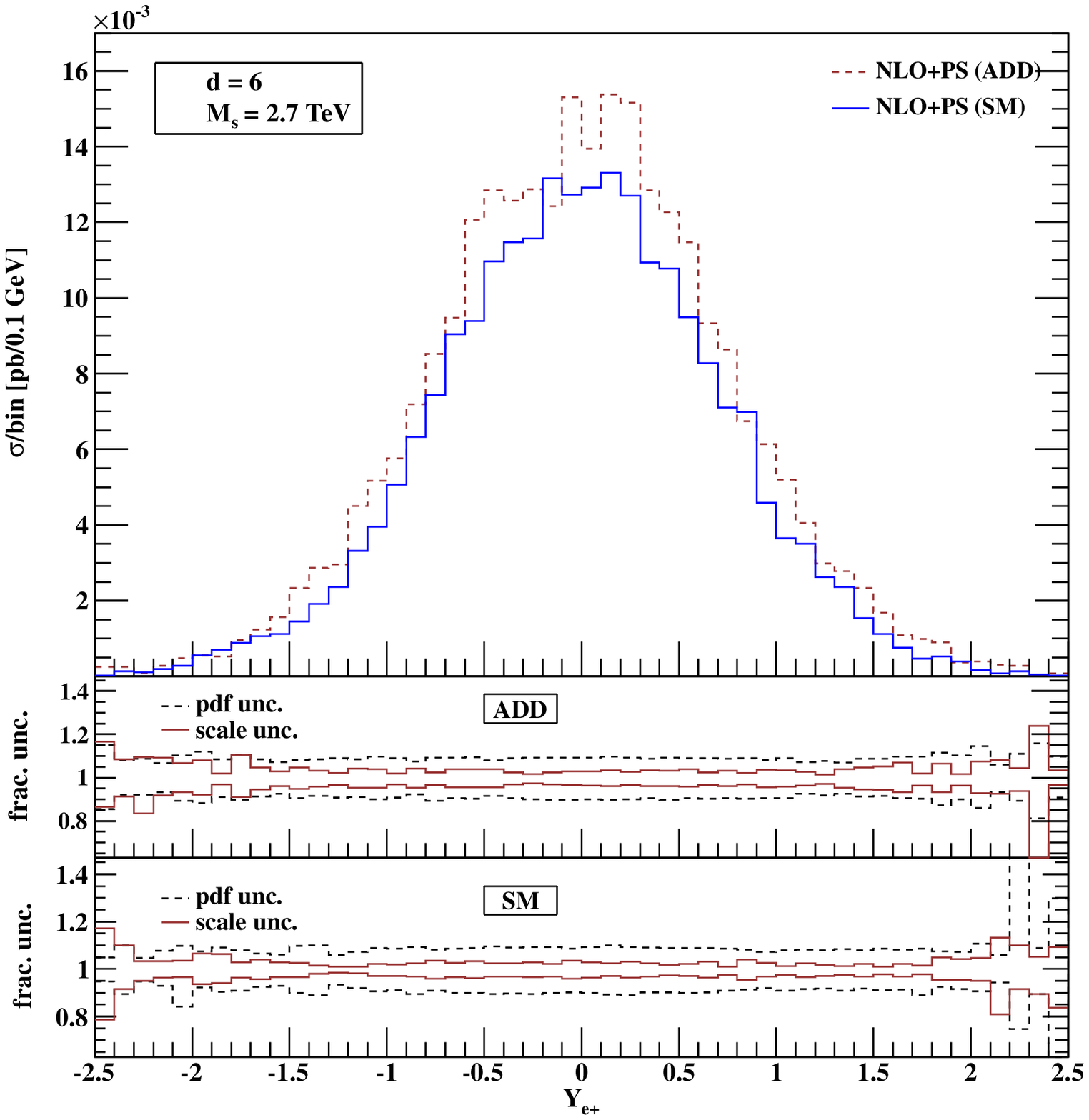}
}
\caption{\label{ww_epptrap}
Transverse momentum distribution (left) and rapidity distribution (right) of the
final state positron which comes from $W^+$ decay for $WW$ production process in
both ADD ($d=6$) and SM when, $2000 < M_{e^+\mu^-\cancel{E_T}} < M_S$ GeV.
}
\end{figure}

For the $WW$ production process, the relevant plots are presented in fig.\
\ref{ww_invmiss} and fig.\ \ref{ww_epptrap}, wherein the decays of $W^\pm$
bosons to leptons and neutrinos are included at the stage of showering.  
For the choice of $M_S$ values associated with specific number of extra
dimensions, we do not find any significant deviation from the SM.  In the
very high invariant mass region of the four-body final state for $d=5,6$
there is some deviation form the SM.  In fig.\ \ref{ww_invmiss}, we have
given the invariant mass ($M_{e^+\mu^-\cancel{E_T}}$) distribution (left)
of the final state decay products of $W^\pm$ and the total missing transverse
energy distribution (right) which comes from the final state neutrinos for
$d=5$. For completeness in fig.\ \ref{ww_epptrap}, we also provide the transverse
momentum distribution of the final state positron (left) along with its rapidity 
distribution (right) for $d=6$.  Only mild difference between the SM and ADD
in the high invariant mass region is observed.  We zoom into this very high
invariant mass region to look for deviations from the SM for these exclusive
observable.  We have studied $d\sigma/d\cancel{E_T}$, $d\sigma/dP_T^{e^+}$
and $d\sigma/d\eta_{e^+}$ in the region when the invariant mass lies between
2 TeV and $M_S$. 


\begin{table}[tbh]
\begin{center}
 \begin{tabular}{|c|c|c|c|c|c|}
  \hline
$d$ &2 &3 &4 &5 &6 \\ \hline
$M_S^{(3{\textcolor{blue}\sigma})}$ (TeV) &12.3 &13.7 &13.5 &11.3 &10.5 \\ \hline
$M_S^{(5{\textcolor{blue} \sigma})}$ (TeV) &10.8 &11.3 &11.1 &11.2 &10.1 \\ \hline
\end{tabular}
\end{center}
\caption{\label{tbl_mslum}
Lower bounds on $M_S$ for various extra dimensions $d$ at the $14$ TeV LHC
with integrated luminosity of $10$ fb$^{-1}$ at 3-sigma and 5-sigma signal
significance.
}
\end{table}
Using the dilepton process, we present the search sensitivity for the extra
dimensions $d=2-6$, for 14 TeV LHC.  The total cross section $\sigma$ is
calculated using the invariant mass distribution of the di-lepton pair for
signal plus background and the background only.  For a particular choice of
extra dimension $d$, we find the minimum luminosity by varying the scale
$M_S$ at 3-sigma $(3 {\textcolor{blue} \sigma})$ and 5-sigma $(5 {\textcolor{blue} \sigma})$ signal significance.
We define the required minimum luminosity as 
$L= max \{ L_{ 3 {\textcolor{blue} \sigma} (5 {\textcolor{blue} \sigma})}, 
L_{ 3{N_S}(5{N_S})}\}$, where 
$L_{ 3 {\textcolor{blue} \sigma} (5 {\textcolor{blue} \sigma)}}$ is the integrated luminosity
at 3-sigma (5-sigma) signal significance and $L_{ 3{N_S}(5{N_S})}$
describes the integrated luminosity to get at least 3(5) signal events.
Now we can get the corresponding $M_S$ value for 10 fb$^{-1}$ luminosity
by inversion which is tabulated in table \ref{tbl_mslum}.  
Of course, a full analysis including the effects of detector simulation,
non-reducible backgrounds {\em etc.}\ can be better performed by the
experimental collaborations.

\section{Conclusion}

The main objective of this work has been to make available, the $\ell^+ \ell^-$,
$ZZ$, $W^+W^-$ production results to NLO+PS accuracy for the large extra dimension
model which is implemented in the {\scriptsize A}MC@NLO framework.  All the
subprocesses that contribute to NLO in QCD have been included for each of these
processes.  A selection of results for 8 TeV LHC has been presented for various
distributions in an attempt to identify region of interest for extra dimension
searches.  Scale and PDF uncertainties for each of these distributions have also
been studied.  In addition, we have presented the search sensitivity for the
extra dimensions $d = 2 - 6$, for 14 TeV LHC at 10 fb$^{-1}$.
With the earlier implementation of the di-photon final state to
the same accuracy \cite{dprecent}, this work completes the rest of the di-final
state process (but for di-jet) in large extra dimension searches.
 In the ADD
model, these codes can be used to generate events of the di-final states
discussed in this paper to NLO+PS accuracy and are available on the website
http://amcatnlo.cern.ch.

\vspace{.5cm}

\noindent
{\large \bf Acknowledgments}
\vspace{.5cm}

The work of MKM and VR has been partially supported by funding from
Regional Center for Accelerator-based Particle Physics (RECAPP),
Department of Atomic Energy, Govt. of India.
We would like to thank the High Performance Computing cluster at Theory
Division, SINP where the computational work was carried out.

\eject


\begin{thebibliography}{99}

\bibitem{cms}
CMS Collaboration, Phys. Rev. Lett. 108 (2012) 111801. 

\bibitem{atlas}
ATLAS Collaboration, Phys.\ Lett.\ B710 (2012) 538.

\bibitem{ADD}
N.\ Arkani-Hamed, S.\ Dimopoulos and G.\ Dvali, Phys.\ Lett.\ B 429 (1998)
263; I.\ Antoniadis, N.\ Arkani-Hamed, S.\ Dimopoulos and G.\ Dvali,
Phys.\ Lett.\ B 436 (1998) 257; N.\ Arkani-Hamed, S.\ Dimopoulos and G.\
Dvali, Phys.\ Rev.\ D59 (1999) 086004.

\bibitem{RS1}
L.\ Randall and R.\ Sundrum, Phys. Rev. Lett. 83 (1999) 3370.

\bibitem{dy}
P.\ Mathews, V.\ Ravindran, K.\ Sridhar and W.\ L.\ van
Neerven, Nucl.\ Phys.\ B713 (2005) 333;
P.\ Mathews, V. Ravindran, Nucl.\ Phys.\ B753 (2006) 1;
M.C. Kumar, P.\ Mathews, V. Ravindran, Eur.\ Phys.\ J.\ C49 (2007) 599.

\bibitem{diph}
M.C. Kumar, P.\ Mathews, V. Ravindran, A.\ Tripathi,
Phys.\ Lett.\ B672 (2009) 45;
%
Nucl.\ Phys.\ B818 (2009) 28.

\bibitem{diZ}
N. Agarwal, V. Ravindran, V. K. Tiwari, and A. Tripathi,
Nucl. Phys. B 830 (2010) 248; Phys.\ Lett.\ B 686 (2010) 244; Phys. Rev. D 82 (2010)
036001.

\bibitem{diW}
N.~Agarwal, V.~Ravindran, V.~K.~Tiwari and A.~Tripathi,
Phys.\ Rev.\ D 82 (2010) 036001; Phys.\ Lett.\ B 690 (2010) 390.

\bibitem{dprecent}
R.~Frederix, M.~K.~Mandal, P.~Mathews, V.~Ravindran, S.~Seth, P.~Torrielli and M.~Zaro,
JHEP 1212 (2012) 102.
  
\bibitem{HLZ}
T.\ Han, J.\ D.\ Lykken and R.\ J.\ Zhang, Phys.\ Rev.\ D59 (1999) 105006.

\bibitem{GRW}
G.\ F.\ Giudice, R.\ Rattazzi, and J.\ D.\ Wells, Nucl.\ Phys.\ B544 (1999) 3.

\bibitem{mcnlo}
S.\ Frixione, B.\ R.\ Webber, JHEP 0206 (2002) 029.

\bibitem{herwig}
G.~Corcella, I.~G.~Knowles, G.~Marchesini, S.~Moretti, K.~Odagiri,
P.~Richardson, M.~H.~Seymour and B.~R.~Webber, JHEP 01 (2001) 010.

\bibitem{madfks}
R.~Frederix, S.~Frixione, F.~Maltoni, T.~Stelzer, JHEP 10 (2009) 003.

\bibitem{fks}
S.~Frixione, Z.~Kunszt, A.~Signer, Nucl. Phys. B467 (1996) 399–442.

\bibitem{mg5}
J.~Alwall, M.~Herquet, F.~Maltoni, O.~Mattelaer, T.~Stelzer, JHEP 1106 (2011) 128.

\bibitem{MadSpin}
P Artoisenet, R Frederix, O Mattelaer, R Rietkerk, 
JHEP 1303 (2013) 015.

\bibitem{mstw}
A.\ D.\ Martin, W.\ J.\ Stirling, R.\ S.\ Thorne, and G.\ Watt,
Eur.\ Phys.\ J.\ C63 (2009) 189–285.


\bibitem{reweight}
R.~Frederix, S.~Frixione, V.~Hirschi, F.~Maltoni, R.~Pittau, et~al.,
JHEP 1202 (2012) 099.

\bibitem{hi_pT}
P. Torrielli and S. Frixione, 
JHEP 1004 (2010) 110.

\end{thebibliography}
\end{document}